\documentclass[letterpaper, 10 pt, conference]{article}  

\usepackage{graphics} 
\usepackage{epsfig} 
\usepackage{amsmath} 
\usepackage{amssymb}  
\usepackage{algorithm,algorithmic}
\usepackage{color}
\usepackage{dsfont}

\newtheorem{remark}{Remark}

\providecommand{\code}[1]{\textbf{\texttt{#1}}} %

\providecommand{\matlab}[0]{\textsc{Matlab}~}
\providecommand{\simulink}[0]{\textsc{Simulink}~}

\providecommand{\ode}[0]{\textbf{ODE}~}

\providecommand{\lti}[0]{\textbf{LTI}~}

\providecommand{\tfirka}[0]{\textbf{TF-IRKA}~}

\providecommand{\ie}[0]{\emph{i.e.}~}
\providecommand{\etc}[0]{\emph{etc.}~}
\providecommand{\eg}[0]{\emph{e.g.}~}

\definecolor{bleuONERA}{RGB}{16,97,169}
\definecolor{grisONERA}{RGB}{64,64,66}

\newenvironment{eq}{\everymath {\displaystyle \everymath{ }} \equation}{ \endequation} %
\renewcommand{\dim}{\mathbf{dim}}
\providecommand{\diag}[0]{\mathbf{diag} } %
\providecommand{\x}[0]{\mathbf{x}} %
\providecommand{\dx}[0]{\mathbf{\dot{x}}} %
\providecommand{\xr}[0]{\mathbf{\hat{x}}} %
\providecommand{\dxr}[0]{\mathbf{\dot{\hat{x}}}} %
\renewcommand{\u}{\mathbf{u}} %
\providecommand{\y}[0]{\mathbf{y}} %
\providecommand{\lv}[0]{\mathbf{l}} %
\providecommand{\rv}[0]{\mathbf{r}} %
\providecommand{\wv}[0]{\mathbf{w}} %
\providecommand{\vv}[0]{\mathbf{v}} %

\providecommand{\Hrealr}[0]{\mathcal{\hat{S}}} %
\providecommand{\Htranr}[0]{\mathbf{\hat{H}}} %
\providecommand{\Er}[0]{{\hat{E}}} %
\providecommand{\Ar}[0]{{\hat{A}}} %
\providecommand{\Br}[0]{{\hat{B}}} %
\providecommand{\Cr}[0]{{\hat{C}}} %
\providecommand{\Dr}[0]{{\hat{D}}} %

\providecommand{\Hreal}[0]{\mathcal{S}} %
\providecommand{\Htran}[0]{\mathbf{H}} %
\providecommand{\E}[0]{{E}} %
\providecommand{\A}[0]{{A}} %
\providecommand{\B}[0]{{B}} %
\providecommand{\C}[0]{{C}} %
\providecommand{\D}[0]{{D}} %

\providecommand{\LL}[0]{{\mathds L}} %
\providecommand{\sLL}[0]{{\mathds L_\sigma}} %

\providecommand{\Htwo}[0]{{\mathcal{H}_{2}}} %
\providecommand{\Hinf}[0]{{\mathcal{H}_{\infty}}} %
\providecommand{\Ltwo}[0]{{\mathcal{L}_{2}}} %
\providecommand{\Linf}[0]{{\mathcal{L}_{\infty}}} %

\providecommand{\Cplx}[0]{\mathbb{C}} %
\providecommand{\Real}[0]{\mathbb{R}} %
\providecommand{\matrixtwo}[4]{ \left[\begin{array}{cc} #1 & #2 \\ #3 & #4 \end{array}\right] } %
\providecommand{\vectortwo}[2]{ \left[\begin{array}{c} #1 \\ #2 \end{array}\right] } %
\providecommand{\vectortwoT}[2]{ \left[\begin{array}{cc} #1 & #2 \end{array}\right] } %
\title{\LARGE \bf
Note on the exact delay stability margin computation of hybrid dynamical systems
}

\author{V. Bellet$^{1}$ and C. Poussot-Vassal and C. Pagetti and T. Loquen$^{2}$
\thanks{*This work was not supported by any organization}
\thanks{$^{1}$Valentine Bellet is with Universit́e de Toulouse, INSA, F-31400 Toulouse, France {\tt\small vbellet@etud.insa-toulouse.fr}}%
\thanks{$^{2}$Charles Poussot Vassal, Claire Pagetti and Thomas Loquen are with ONERA - The French Aerospace Lab, F-31055 Toulouse France
        {\tt\small Charles.Poussot-Vassal@onera.fr, Claire.Pagetti@onera.fr, Thomas.Loquen@onera.fr}}%
}

\begin{document}

\maketitle
\thispagestyle{empty}
\pagestyle{empty}

\begin{abstract}
Traditionally, the delay margin of a looped system is computed by considering both the controller and system representations that evolve in the same space (\eg either continuous or discrete-time). However, as in practice the system is continuous and the controller is mostly embedded in a computer, the looped - controller / system pair - model is hybrid. As a consequence, the computed delay margin might vary with respect to the continuous (or discrete one). This paper proposes a novel approach to compute the exact delay margin of hybrid systems, and more specifically, when a discrete-time controller is looped with a continuous-time system. The main interest is then to provide the practitioners with a way to select the appropriate discretization technique for maximizing the delay margin and to be able to exactly evaluate the delay margin before implementation on target. The main idea is to approximate the discrete-time controller with an equivalent continuous-time one (often with higher order) and to exploit the classical continuous-time frequency-based analysis strategies.  
\end{abstract}

\section{Introduction}
\label{sec:intro}

\subsection{Foreword and preliminary comments}

Most of the engineering systems or processes are naturally modeled using a set of continuous-time linear ordinary differential and algebraic equations $\Hreal$ as 
\begin{eq}
\Hreal : \left\{
\begin{array}{rcl}
\E\dx(t) &=& \A\x(t)+\B\u(t)\\
\y(t) &=& \C\x(t)+\D\u(t)
\end{array}
\right. 
\label{eq-ltiS}
\end{eq}
which associated complex-valued $n_u\times n_y$ transfer function reads $\Htran(s)=\C(s\E-\A)^{-1}\B+\D \in \Linf$, where $\Linf$ stands for the space of (here rational) meromorphic functions with no singularities on the imaginary axis ($n_u$ and $n_y$ stand for the input and output dimensions). Then, based on $\Hreal$ as given in \eqref{eq-ltiS}, a stabilizing and (generally) stable linear continuous-time controller equipped with a realization
\begin{eq}
\Hreal_c : \left\{
\begin{array}{rcl}
\dx_c(t) &=& \A_c\x_c(t)+\B_c\y(t)\\
\u(t) &=& \C_c\x(t)+\D_c\y(t)
\end{array}
\right. 
\label{eq-ltiCont}
\end{eq}
with $n_y\times n_u$ complex-valued transfer function $\Htran_c(s)=\C_c(sI-\A_c)^{-1}\B_c+\D_c \in \Hinf$, is designed. Such a control law is traditionally optimized using any continuous-time controller tuning method to ensure closed-loop stability and some performances. Then, stability margins, such as the delay one (\textbf{DM}), defined as the maximal delay in the loop leading to closed-loop instability, are usually measured before further implementation. However, even if controllers are mostly designed in the continuous-time domain, the associated control law is almost always implemented on a computer working with a constant sampling time $h\in\Real_+$, involving a discrete-time representation $\Hreal_d$ of the controller \eqref{eq-ltiCont}, described as a set of difference equations as \cite{Astrom:1997, Wittenmark_book_2002}
\begin{eq}
\Hreal_d : \left\{
\begin{array}{rcl}
\x_d(k+h) &=& \A_d\x_d(k)+\B_d\y(k)\\
\u(k) &=& \C_d\x_d(k)+\D_d\y(k)
\end{array}
\right. 
\label{eq-ltiDisc}
\end{eq}
with $n_y\times n_u$ complex-valued transfer function $\Htran_d(z)=\C_d(zI-\A_d)^{-1}\B_d+\D_d \in \mathbb D$, where $\mathbb D$ stands for the unit disk centered in 0.  With reference to \eqref{eq-ltiDisc}, $k$ denotes the sampling index. Moreover, for sake of simplicity, as we considered $\Hreal_c$ to be a stable (\ie $\in \Hinf$) control dynamical process, we will also consider $\Hreal_d$ as a stable one (\ie $\in \mathbb D$)\footnote{Still, reader may consider unstable controllers. However, this will weight the presentation of the present paper.}. 

When a continuous dynamical operator, here the controller $\Hreal_c$, is given as a transfer function $\Htran_c\in\Hinf$, to obtain the exact discrete representation $\Hreal_d$, the Laplace differential operator $s$ must be replaced by an (irrational) difference operator $z=e^{sh}$ (thus $s=\frac{1}{h}\ln(z)$ ), where $h$ is the sampling time. Such an operation leads then to an irrational complex-valued transfer function $\Htran_{irrat.}$ given as
\begin{eq}
\Htran_{irrat.}(z)=\C_c\bigg(\frac{1}{h}\ln(z)I-\A_c\bigg)^{-1}\B_c+\D_c
\label{eq-irrat}
\end{eq}
In practice, such an irrational operator is approximated with a rational form according to the selected implementation scheme. This approximation is usually based on different exponential series expansion. To this aim, several algorithms with advantages and drawbacks can be used to approximate the continuous controller as a set of difference equations. This problem has been addressed for many years and there exist methods involving dedicated criteria such as \cite{Cantoni_tac_2004, KUO1973, SAWADA2004}. In practice, the classical methods known as the backward, forward or bilinear are used. These latter lead to different values of the $\{I,\A_d,\B_d,\C_d,\D_d,h\}$ quadruplet defining the so-called discrete-time model given in \eqref{eq-ltiDisc}. In such a case, the main concern will be the conservation of the controller stability (or eventually instability). Then, hopefully the performances and delay margin computed using the continuous system \eqref{eq-ltiS} and controller \eqref{eq-ltiCont} are kept the same as $h$ is small enough. However, the discretization step always introduces modifications and signal distortion.


Although techniques for directly synthesizing discrete-time control laws are available in the literature (see \eg \cite{DTC_2015, Jungers_IJC_2016}), the most common practice for control engineers,  is to design first a continuous-time controller, which is afterwards converted into a discrete one. Several considerations lead to this choice: 
\begin{itemize}
\item Direct design of discrete-time controllers requires a pre-determination of the sampling time in order to model plant, noise, process, disturbances, \etc \cite{keller_90}. Upper bounds for the sampling time can only be obtained after considering closed-loop bandwidth and, depending on specifications, this may only be available after controller design. Since this obvious dilemma does not occur in continuous-time controller design, it may be reasonable to design first a continuous-time controller, which is in a second step approximated by a discrete-time one. 


\item Due to the continuous nature of many dynamical systems, the design of continuous-time controller is easiest and seems more appreciated by the practitioners (researchers may also note that the control literature is way larger). Engineers are naturally more comfortable with continuous-time methods (manual tuning of controller's gains linked to physical parameters of the plant) and lot of efficient design methods are difficult to transpose to the discrete case.
\end{itemize}

Moreover, one may note that discretizing first the model, then synthesizing a control law directly in the discrete-time domain, would lead to the same problem of stability analysis of hybrid system when looping with the real continuous-time model.

In all cases, the discrete-time controller $\Hreal_d$ given in \eqref{eq-ltiDisc} is an approximation of the nominal continuous-time one. Therefore, one may expect an impact on the behavior of the closed-loop system when connecting the actual plant and the embedded controller. 

To deal with this problem, delays  introduced in the actual control loop due to the corresponding embedded architecture (communication process, scheduling and computation tasks) \cite{ALBERTOS2005, Naghshtabrizi2010ImplementationCF} have also to be considered. Here again, the stability and performance of dynamical system including delays, jitters or drop-outs is addressed in the literature. However, the actual interconnection discrete-time controller / continuous-time plant is rarely taken into account and to the best of our knowledge, classical stability or performance criteria cannot easily be derived.  Note however that there exist numerical tools  allowing the simulation of  embedded controllers (real-time implantation, scheduling, network transmissions, ...) and continuous plant dynamics like \matlab \simulink toolboxes \cite{cervin03, Cremona_tres_2015}, Ptolemy-HLA distributed co-simulation framework \cite{LasnierCSPD13} and approaches involving hardware in the loop. 

Thus, when implementating a controller in a real digital computer architecture, the following question should first arises: how the digital controller will impact stability and performance of the closed-loop system ? Behind this question, there is the issue of stability and performance analysis of hybrid systems, which is a difficult point, in spite of recent breakthrough \cite{HDS_2012}. 
	
\subsection{Contribution}

Given the above considerations, in this paper, we propose to compute a more exact delay margin (denoted \textbf{DM}), taking explicitly the connection of the continuous-time plant and discrete-time controller. The proposed solution takes benefit from the use of advanced model approximation methods. More specifically, by exploiting recent improvement in dynamical system approximation such as the data-driven Loewner framework (see \eg \cite{AntoulasSurvey:2016} and \cite{MeyerIFASD:2017,PoussotMATHMOD:2018} for some applications) or $\Htwo$-oriented approximation techniques embedded in the \tfirka (see \eg \cite{Beattie:2009,Beattie:2016} and \cite{DalmasECC:2016} for applications), this paper proposes a novel method to compute the exact delay margin. The idea, detailed and illustrated in the rest of the paper, consists in approximating the exact discrete-time complex-valued irrational model of the controller $\Htran_{irrat.}(z)$ given as in \eqref{eq-irrat} by a continuous-time complex-valued rational function $\Htranr_d(s)$, which can be used in place of \eqref{eq-irrat} for evaluating the classical continuous-time delay margin computation methods. The overall approach is validated on a simple example, illustrating the difference between the continuous theoretical and real hybrid delay margin.

\begin{remark}[About the ``real'' delay margin]
Along the paper, we talk about ``real'' delay margin. This terminology might be complex to define in the context of hybrid systems. Here, we will consider the real delay margin using two different approaches: the \emph{(i)} \ode approach and the \emph{(ii)} real-time software approach. First, in the former case \emph{(i)}, the minimal delay leading to the instability is obtained when solving the looped continuous / discrete-time system with an \ode solver and an adaptive step. As a matter of consequence, we will consider a destabilizing case when the \ode solver leads to a diverging output signal sequence. Obviously, from a numerical point of view, this also might be questionable. Second, when considering the latter case \emph{(ii)}, a dedicated real-time environment, different from the traditional \matlab based one, is being used to cross check the approach (this latter is detailed at the end of the paper).
\end{remark}



\subsection{Notations and paper structure}

\textbf{Notations:} in this paper we denote $\mathbb N$, $\Real$ and $\Cplx$, the natural, real and complex values sets. The $\Htwo$ space denotes the set of complex-valued matrix functions $\Htran(s)$ analytic over $\Cplx_+$ and which inner product integral is bounded along the imaginary axis, the $\Hinf$ space denotes the ones which supremum along the imaginary axis is bounded. In addition, $\mathcal{RH}_\bullet$ denotes the spaces of rational functions in $\mathcal{H}_\bullet$. Then $\Linf(\imath\Real)=\Hinf(\Cplx_+) \oplus\Hinf(\Cplx_-)$ and $\Ltwo(\imath\Real)=\Htwo(\Cplx_+) \oplus\Htwo(\Cplx_-)$. $\Htran(s)$ denotes the complex-valued transfer function, where $s$ is the Laplace variable, and $\Hreal=(\E,\A,\B,\C,\D)$ its associated realization of dimension $n$. Similarly, the discrete-time transfer function, sampled at period $h\in \Real_+$, $\Htran(z)$ denotes the complex-valued transfer function, where $z$ is the $Z$-transform variable, and $\Hreal=(\E,\A,\B,\C,\D,h)$. Then, $\mathbb D$ and $\overline{\mathbb D}$ denote the unit circle and its complement, respectively. 

\textbf{Structure of the paper:} after introduction in Section \ref{sec:intro}, Section \ref{sec:result} describes the main result of the paper, \ie a methodology to obtain the exact delay margin of a linear hybrid dynamical system, as well as a numerically reliable procedure, denoted as Hybrid Systems Delay Margin Algorithm (\textbf{HSDMA}). A numerical example, illustrating the proposed approach is given in Section \ref{sec:example}. Conclusions and perspectives are discussed in Section \ref{sec:conclusion}.



\section{Main result: approximation-based hybrid systems delay margin computation}
\label{sec:result}

\subsection{Algorithm general idea}

Let us first describe the proposed Hybrid Systems Delay Margin Algorithm (\textbf{HSDMA}) as in Algorithm \ref{algo}. The steps of the following algorithm are detailed after.

\begin{algorithm}
\caption{Hybrid Systems Delay Margin Algorithm (\textbf{HSDMA})}
\begin{algorithmic}[1]
\REQUIRE $\Htran(s)$ as in \eqref{eq-ltiS}, $\Htran_d(z)$ as in \eqref{eq-ltiDisc}
\STATE Set $N\in\mathbb N$ s.t. $N\gg \dim(\A_d)$
\STATE Set frequency grid $\{\omega_i\}_{i=1}^N \in \Real_+$ satisfying
\begin{eq}
0<\omega_i \leq\pi/h
\end{eq}
\STATE Compute the frequency response of $\Htran_d(z)$ \eqref{eq-ltiDisc} over $\{\omega_i\}_{i=1}^N$ as
\begin{eq}
\{\Phi_i\}_{i=1}^N = \Htran_d(e^{\imath \omega_i h})
\end{eq}
\STATE Given $\{e^{\imath \omega_i h},\Phi_i\}_{i=1}^N$, compute the approximate model $\Htranr_d\in\Hinf$ or $\in \Linf$ satisfying, for $i=1,\dots,N$
\begin{eq}
\Htranr_d(\imath\omega_i) = \Phi_i
\label{eq-interp}
\end{eq}
\STATE[opt.] If $\Htran_d(z)$ is stable (\ie $\in \mathbb D$), enforce $\Htranr_d(s)$ to be stable too (\ie $\in \Hinf$)
\STATE Compute $\mathbf L(s)=\Htranr_d(s)\Htran(s)$
\STATE Compute the delay margin \textbf{DM}, based on $\mathbf L(s)$
\end{algorithmic}
\label{algo}
\end{algorithm}

With reference to Algorithm \ref{algo}, the \textbf{HSDMA} procedure first samples the frequency response of the discrete-time controller $\Htran_d(z)$ up to the Nyquist frequency defined by $\pi/h$. Indeed, it is pointless to sample over with frequency since the spectrum is simply repeated (Steps 1-3). Once this step is achieved, input-output data of the frequency response of the discrete-time controller are available and denoted as $\{\omega_i,\Phi_i\}_{i=1}^N$. Then, based on this data set, one then computes  a rational interpolant model $\Htranr_d$ that ensures $\Htranr_d(\imath\omega_i) = \Phi_i$ (Step 4). This step may be achieved by any method, but here, we suggest to employ the Loewner framework \cite{Mayo:2007}, recalled in the next subsection. Indeed, one main strength of this method is that   it enables ensuring interpolatory conditions as in \eqref{eq-interp} with the rational model embedding the minimal McMillian degree $r$ (this latter being numerically computed using the Loewner pencil). Therefore, if $N$ is sufficiently large (\ie $ N\gg r$), the continuous-time model $\Htranr_d(s)$ will perfectly reproduce the discrete-time one $\Htran_d(z)$, over the considered interval (up to $\pi/h$). Note that the case where under-sampling has been selected, \ie if $r=N$, one may simply re-sample with an increased $N$. Then, once $\Htranr_r(s)$, the continuous-time rational approximation of order $r$ is obtained, one may simply compute the delay margin as traditionally done with any continuous-time delay margin method (Steps 6-7). Note that if the original controller  $\Htran_d(z)$ is stable, \ie eigenvalues lie inside the unit disk $\mathbb D$, the approximated continuous-time rational controller $\Htranr_d(s)$ should also be stable, \ie eigenvalues  should lie in $\Cplx_-$.

The optional step 5 may be achieved following the proposed approach given in \cite{Kohler:2014}, let to the reader discretion.

\subsection{Notes on the Loewner-based approach for discrete-time $\Htran_d(z)$ rational approximation with continuous-time $\Htranr_d(s)$}

Here we detail how to complete steps 3-4 of Algorithm \ref{algo}, when a discrete-time controller $\Htran_d(z)$ is given. After sampling of $\Htran_d(z)$ at $z=e^{\imath \omega_ih}$, one obtains the frequency responses $\{\Phi_i\}_{i=1}^N$ and thus the couple, 
\begin{eq}
\{e^{\imath \omega_ih}, \Phi_i \}_{i=1}^N
\label{eq:dataFreq}
\end{eq}
a set of frequency-domain (or complex-domain) $n_y$ inputs, $n_u$ outputs data $\Phi_i\in \Cplx^{n_u\times n_y}$ collected at varying frequencies $e^{\imath \omega_ih}\in \Cplx$ obtained by numerical simulation. These data satisfy, for $i=1,\dots,N$,
\begin{eq}
\overline{\u}(e^{\imath \omega_ih}) = \Phi_i \overline{\y}(e^{\imath \omega_ih})
\end{eq}
where $\overline{\u}(s)\in\Cplx^{n_u}$, $\overline{\y}(s)\in\Cplx^{n_y}$ respectively are the Fourrier transform values of the inputs $\u(t) \in \Real^{n_u}$ and outputs $\y(t) \in \Real^{n_y}$, evaluated at $e^{\imath \omega_ih}$. In this setting,  the objective is to find a \lti dynamical model $\Htranr_d(s)=\Cr(s\Er-\Ar)^{-1}\Br+\Dr$, equipped with a realization of ``complexity'' $r\in\mathbb N$  given as
\begin{eq}
\Hreal: \left\{ 
\begin{array}{rcl} 
\Er\dxr(t) &=&\Ar \xr(t) +\Br \y(t)  \\ 
\u(t) &=& \Cr \xr(t) + \Dr \y(t)
\end{array} \right. 
\label{eq:Hreal}
\end{eq}
where $\xr(t) \in \Real^{r}$ denotes the internal variables\footnote{$\xr(t) \in \Real^{r}$ are the state variables if $\Er$ is invertible.} 
 and where 
\begin{eq}
\Er,\Ar \in \mathbb{R}^{r \times r}, \Br \in \mathbb{R}^{r\times n_y}, \Cr \in \mathbb{R}^{n_u \times r} \text{ and } \Dr \in \mathbb{R}^{n_u \times n_y} \nonumber
\end{eq}
are constant matrices, for which the frequency response $\Htranr_d(e^{\imath \omega_ih})$ well reproduces the data set $\{e^{\imath \omega_ih},\Phi_i\}_{i=1}^N$.

To this aim, the Loewner matrices offer a versatile and compliant framework to deal with frequency-domain data, by seeking for a realization interpolating these data, in the barycentric sense. More specifically, let be given left interpolation driving frequencies $\{\mu_j\}_{j=1}^q \in \Cplx$ with left output or tangential directions $\{\lv_j\}_{j=1}^q \in \Cplx^{n_u}$, producing the left responses $\{\vv_j\}_{j=1}^q \in \Cplx^{n_y}$ and right interpolation driving frequencies $\{\lambda_i\}_{i=1}^k \in \Cplx$ with right input or tangential directions $\{\rv_i\}_{i=1}^k \in \Cplx^{n_y}$, producing the right responses $\{\wv_i\}_{i=1}^k\in \Cplx^{n_u}$, one aims at finding a realization $\Hrealr$ such that the resulting transfer function $\Htranr_d$ is a tangential interpolant of the data, \ie satisfies the following left and right interpolation conditions (note that here $N=q+k$ and $\{e^{\imath \omega_ih}\}_{i=1}^N=\{\mu_j\}_{j=1}^q\bigcup\{\lambda_i\}_{i=1}^k$):
\begin{eq}
\left.
\begin{array}{c}
\lv_j^*\Htranr_d(\mu_j) = \vv_j^* \\
\text{for $j=1,\dots, q$}
\end{array}
\right\}
\text{~~and~~}
\left\{
\begin{array}{c}
\Htranr_d(\lambda_i)\rv_i = \wv_i \\
\text{for $i=1,\dots, k$}
\end{array}
\right. 
\label{eq:interpData}
\end{eq}
In practice, $\mu_j$ and $\lambda_i$ are subsets of $\{e^{\imath \omega_ih}\}_{i=1}^N$, $\vv_j$ and $\wv_i$ are subsets of $\Phi_i$. Tangential directions are arbitrarily chosen in the general case, although it is important  ensuring that no rank loss are observed in the Sylvester equations, later detailed \eqref{eq:sylv1}-\eqref{eq:sylv2}, leading to transfer matching inaccuracy.

The main ingredient to  achieve \eqref{eq:interpData} is the Loewner matrix, which was developed in a series of papers (see \eg \cite{Mayo:2007,Ionita:2014,AntoulasSurvey:2016}). In the sequel, we recall the main steps. Let be given, the left or row data and the right or column data:
\begin{eq}
(\mu_j,\lv_j^*,\vv_j^*), j=1,\dots ,q \text{~~and~~} (\lambda_i,\rv_i,\wv_i), i=1,\dots ,k
\label{eq:data}
\end{eq}
Moreover, let us assume that $\lambda_i$ and $\mu_j$ are distinct, then the associated Loewner $\LL \in \Cplx^{q\times k}$ and shifted Loewner $\sLL \in \Cplx^{q\times k}$ matrices, also referred to as the Loewner pencil, are constructed as follows, for $i=1,\dots,k$ and $j=1,\dots,q$:
\begin{equation}
[\LL]_{j,i} = \dfrac{\vv_j^*\rv_i - \lv_j^*\wv_i}{\mu_j - \lambda_i} \mbox{~~,~~}
\,[\sLL]_{j,i} = \dfrac{\mu_j\vv_j^*\rv_i - \lambda_i\lv_j^*\wv_i}{\mu_j - \lambda_i}.
\label{eq:LL}
\end{equation}
Then, by organizing the left and right interpolation data as:
\begin{eq}
\left.
\begin{array}{rcl}
\mathbf M &=&  \diag (\mu_1,\dots,\mu_q) \in \Cplx^{q\times q}\\
\mathbf L^* &=&  [\lv_1 ~\dots ~ \lv_q] \in \Cplx^{n_u\times q}\\
\mathbf V^* &=&  [\vv_1 ~\dots ~ \vv_q] \in \Cplx^{n_y\times q}
\end{array}
\right\}
\label{eq:interpLeftMatrix}
\end{eq}
and
\begin{eq}
\left.
\begin{array}{rcl}
\mathbf \Lambda &=&  \diag (\lambda_1,\dots,\lambda_k) \in \Cplx^{k\times k}\\
\mathbf R &=&  [\rv_1 ~\dots ~ \rv_k] \in \Cplx^{n_y\times k}\\
\mathbf W &=&  [\wv_1 ~\dots ~ \wv_k] \in \Cplx^{n_u\times k}
\end{array}
\right\}
\label{eq:interpRightMatrix}
\end{eq}
the Loewner $\LL$ and shifted Loewner $\sLL$ matrices \eqref{eq:LL} satisfy the following Sylvester equations:
\begin{eq}
\LL \mathbf \Lambda - \mathbf M \LL  =  \mathbf L \mathbf W - \mathbf V \mathbf R
\label{eq:sylv1}
\end{eq}
and 
\begin{eq}
\sLL \mathbf \Lambda - \mathbf M  \sLL = \mathbf L \mathbf W \mathbf \Lambda - \mathbf  M \mathbf V \mathbf R
\label{eq:sylv2}
\end{eq}

Consequently, following the main results of \cite{Mayo:2007}, given the right and left interpolation data  as in \eqref{eq:data}, and assuming that $k = q$,  $(\LL,\sLL)$ is a regular pencil where $\lambda_i$ or $\mu_j$ are not eigenvalues, and which has been simplified using any rank revealing factorization such as 
\begin{eq}
\begin{array}{rcl}
\LL&\leftarrow& -\mathbf Y_1^*\LL \mathbf X_1 \\ 
\sLL&\leftarrow& -\mathbf Y_1^*\sLL \mathbf X_1 \\
\mathbf V&\leftarrow&\mathbf Y_1^*\mathbf V \\
\mathbf W&\leftarrow&\mathbf W \mathbf X_1
\end{array}
\end{eq}
where 
\begin{eq}
\LL=\vectortwoT{\mathbf Y_1}{\mathbf Y_2} \matrixtwo{\Sigma_1}{}{}{\Sigma_2}\vectortwo{\mathbf X_1^*}{\mathbf X_2^*},
\end{eq}
the rational transfer function $\Htranr_d(s)=\Cr(s\Er-\Ar)^{-1}\Br$, with realization $\Hrealr : (\Er,\Ar,\Br,\Cr,0)$ constructed as
\begin{eq}
\Er= -\LL, \ \Ar= -\sLL, \ \Br=\mathbf V\ \text{~~and~~} \ \Cr=\mathbf W
\end{eq}
is a minimal descriptor realization and 
\begin{eq}
\Htran(s)= \mathbf W(\sLL - s\LL)^{-1}\mathbf V
\end{eq}
interpolates the left and right constraints, \ie ensures \eqref{eq:interpData}. 

\subsection{Theoretical considerations and comments}

Now the main procedure and Loewner framework, tailored to approximate a discrete-time dynamical model by a continuous-time one, has been detailed, let us provide some arguments to assess the soundness of the approach given in Algorithm \ref{algo}.
 
The main idea behind the proposed procedures lies on the fact that the delay margin is a frequency-based criteria which considers a Nyquist-like criteria. Therefore, if one is able to accurately approximate any discrete-time controller $\Htran_d(z)$ by a rational continuous-time one $\Htranr_d(s)$, then the standard delay margin may be computed simply in the continuous-time domain. The trick then stands in being able to generate $\Htranr_d(s)$. Actually, as above exposed, the Loewner framework facilitates that by providing the minimal realization matching $\Htran_d(z)$. In practice, this may lead to dimension of $\Htranr_d(s)$ greater than $\Htran_d(z)$. Indeed, as illustrated in the example Section \ref{sec:example}, the discretization of a continuous controller often introduces some distortion of the signal phase, which may only be captured with irrational term or a larger realization.


\section{Numerical application}
\label{sec:example}

\subsection{Application of the \textbf{HSDMA}}

As the overall procedure has been detailed in Section \ref{sec:result}, let us illustrate it on a simple but yet representative example, by focusing, for didactic reasons, on the engineering soundness. Let us consider \code{P} a single input - single output plant defined its transfer function or a state-space realization as in \eqref{eq-ltiS} given as:
\begin{eq}
\code{P}:\left\{
\begin{array}{rcl}
\dx &=& \begin{bmatrix}-10 &-5 \\ 4& 0\end{bmatrix} \x + \begin{bmatrix} 0.5 \\ 0\end{bmatrix} \u\\
\y &=& \begin{bmatrix}0 & 0.5\end{bmatrix}\x 
\end{array} 
\right. 
\end{eq}

\begin{figure}[!htb]
\begin{center}
\includegraphics[width=\columnwidth]{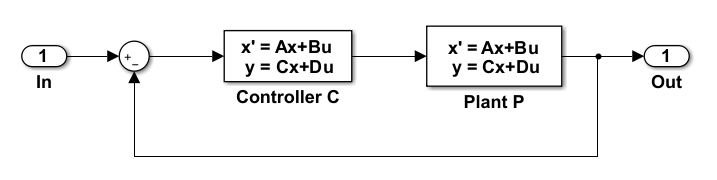}
\caption{Continuous-time closed-loop system}
\label{fig:bf_cont}
\end{center}
\end{figure}

In addition, let us suppose that a continuous-time PI-like controller \code{C}, with realization \eqref{eq-ltiCont}, has been designed such that the output feedback interconnection, as represented Figure \ref{fig:bf_cont}, is asymptotically stable. Moreover, the controller \code{C} is designed to be Hurwitz. Then it reads
\begin{eq}
\code{C}:\left\{
\begin{array}{rcl}
\dx_c(t) &=& \begin{bmatrix} -0.001 & 7,854 \\ 0 &-62.83\end{bmatrix} \x(t) + \begin{bmatrix}0 \\ 8\end{bmatrix} \y(t)\\
\u(t) &=& \begin{bmatrix} 70 & 235.6\end{bmatrix} \x(t)
\end{array} 
\right. 
\end{eq}

In this negative feedback control architecture, one can easily compute the delay margin \eg using the \matlab function \code{allmargin}. In this case, one obtains a delay margin of $0.3254$ seconds. Now by introducing a transport delay block in the above  \matlab \simulink environment, using and \ode solver with adaptive time step, one can observe that a delay of $0.3254$ seconds leads to an auto-oscillating output (stable before, and unstable after).

\begin{figure}[!htb]
\begin{center}
\includegraphics[width=\columnwidth]{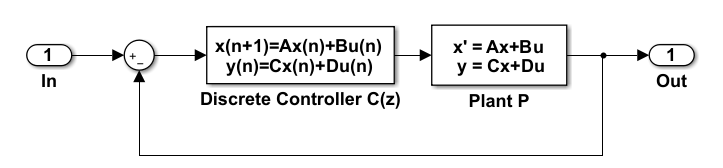}
\caption{Discrete-time closed-loop system}
\label{fig:bf_d}
\end{center}
\end{figure}

In view of its implementation, controller \code{C} is now discretized with a constant sampling time $h$. The resulting hybrid closed-loop is now represented on Figure \ref{fig:bf_d}, where the continuous-time controller has been replaced by its discrete-time version. 

\begin{figure}[H]
\begin{center}
\includegraphics[width=\columnwidth]{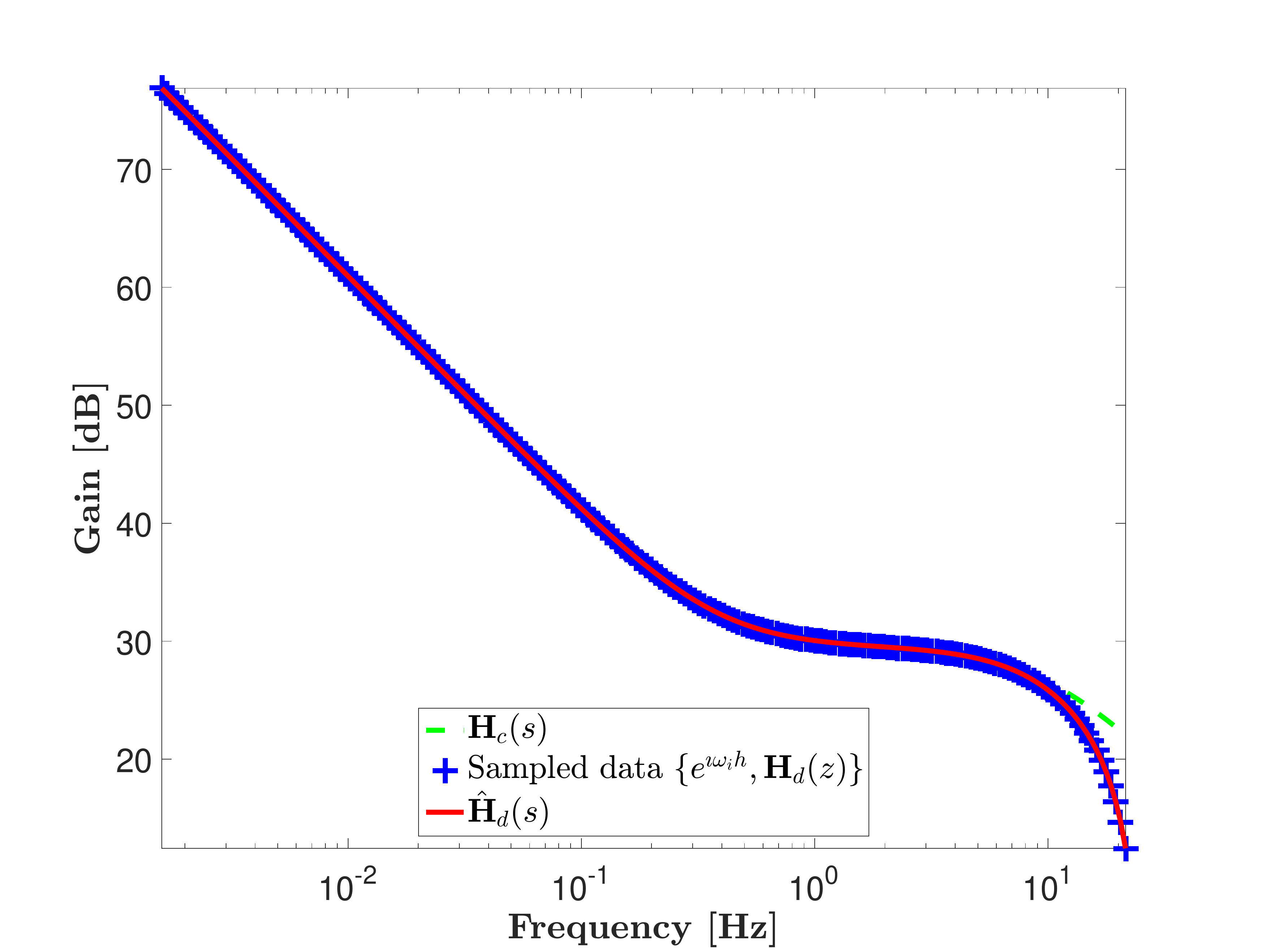}
\includegraphics[width=\columnwidth]{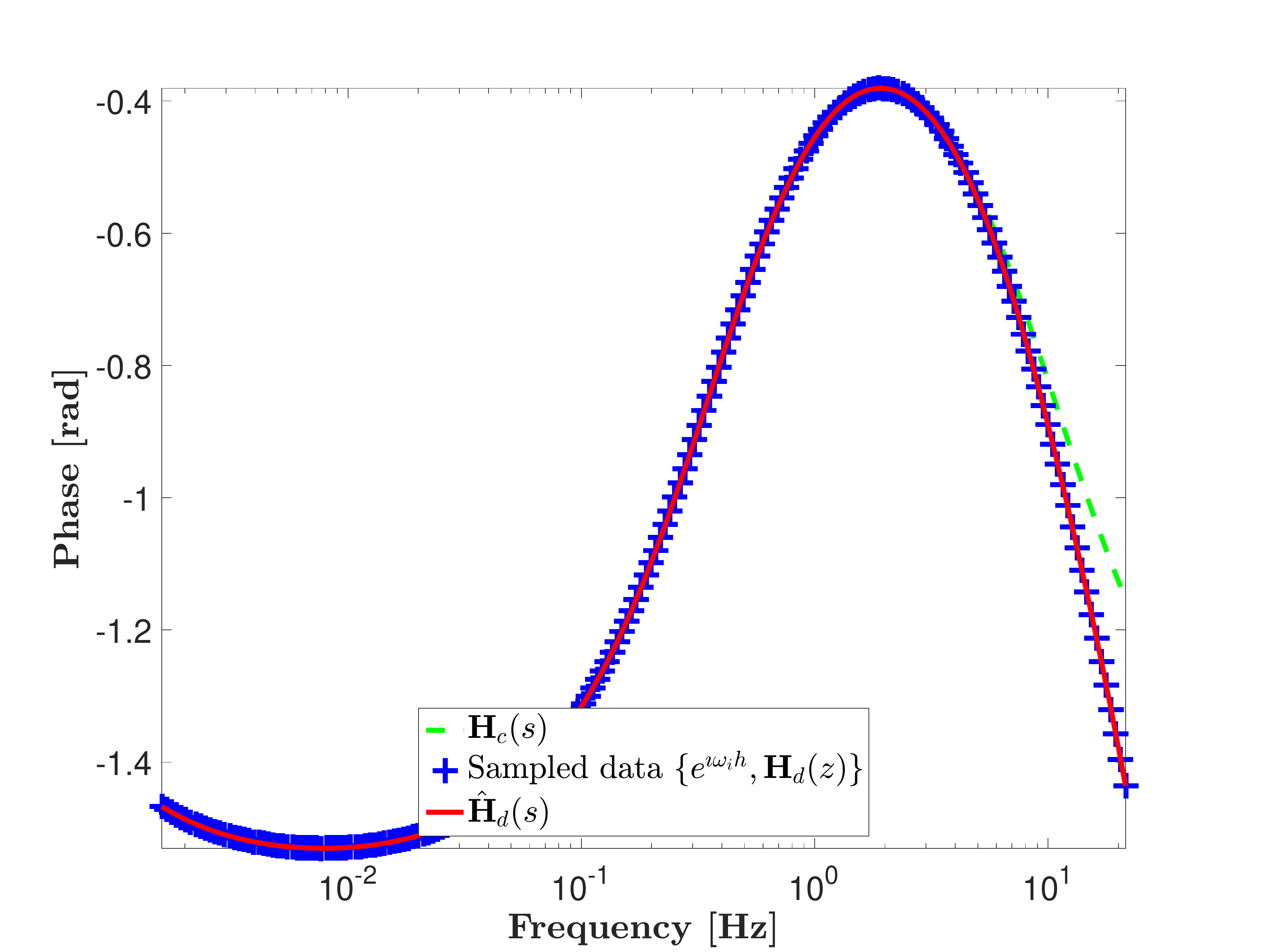}
\caption{Controller Bode sampled gain (top) phase (bottom) responses of the original continuous controller $\Htran_c(c)$, discrete-time $\Htran_d(z)$ (blue crosses), and its approximation $\Htranr_d(s)$, obtained at step 4 of \textbf{HSDMA}.}
\label{fig:exple01_bode}
\end{center}
\end{figure}

By selecting $h = 0.02$ and a bilinear approximation method to get $\Htran_d(z)$, one can easily compute $\Htranr_d(s)$ thanks to \textbf{HSDMA}. This new controller, obtained for $N=200$ logarithmically spaced frequency values, reproducing exactly the  representation $\Htran_d(z)$ has $r=18$ state variables and the poles are between $-394.04 \pm 876.74i$ and $0$. It perfectly fits the behavior of the discrete one as $\Htran_d(z)$ (the model is not given here for space limitations but Bode responses are plotted on Figure \ref{fig:exple01_bode}). 

Thanks to this continuous controller, one can now compute the delay margin \eg with \code{allmargin}, proving a result $0.3255$ which is a little bit higher than with the original continuous controller (which was $0.3254$). Indeed, this difference comes from the small differences close to the Nyquist frequency in the Bode response between $\Htran_c(s)$ and $\Htran_d(z)$.

\begin{figure}[H]
\begin{center}
\includegraphics[width=\columnwidth]{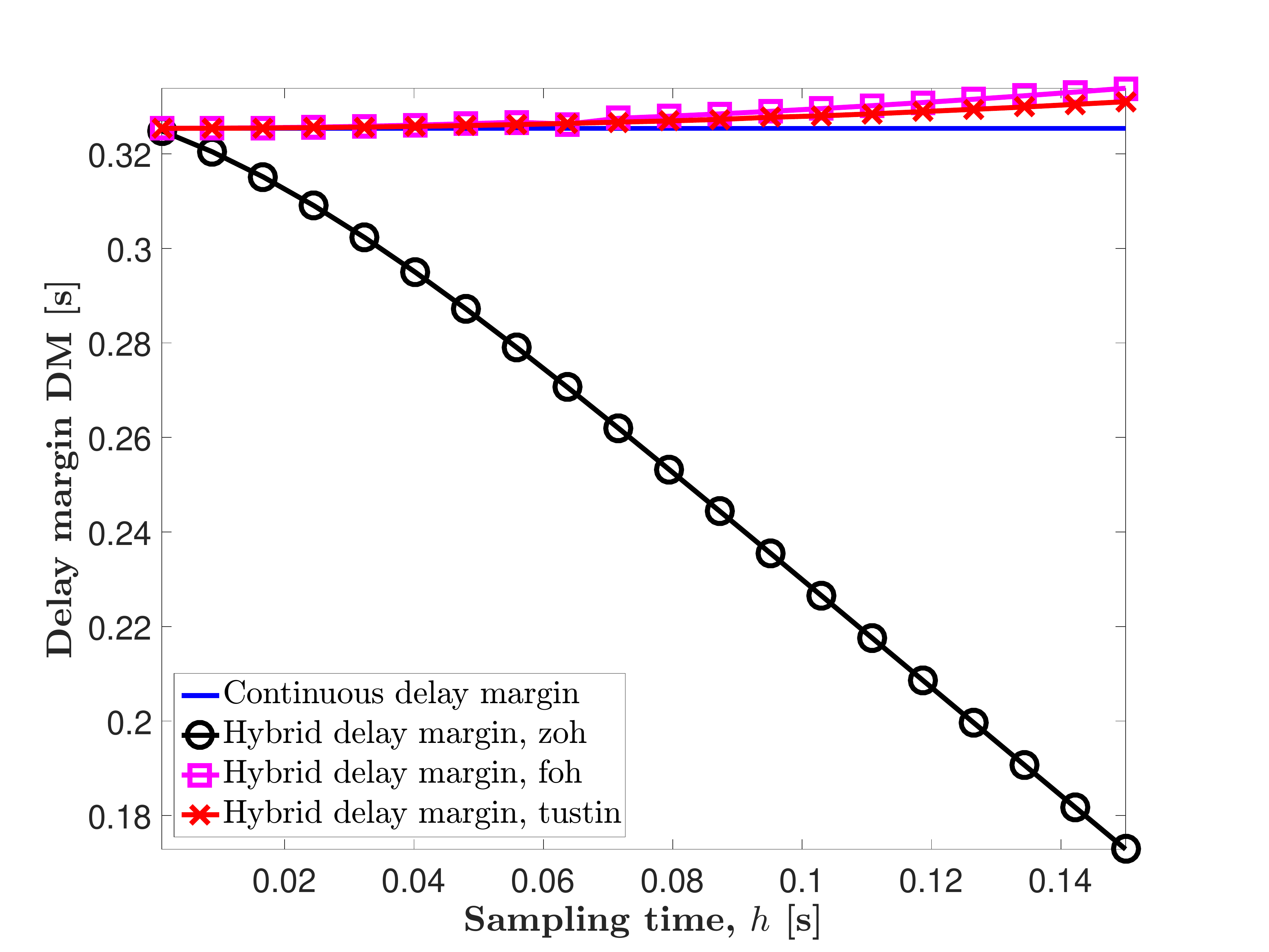}
\caption{Delay margin obtained when considering $\Htran(s)$ interconnected to $\Htran_c(s)$ (solid blue), $\Htranr_d(s)$ obtained with backward (black rounded), $\Htranr_d(s)$ obtained with forward (pink squared) and $\Htranr_d(s)$ obtained with bilinear (red crossed) transform.}
\label{fig:exple01_margin}
\end{center}
\end{figure}

Still using this illustrative example, let us now consider that the controller has been sampled with different periods $0<h\leq0.15$ in order to catch the continuous-time closed-loop bandwidth (which is close to $5.7$ rad/sec).  Moreover, let us consider different discretization methods, namely, backward, forward and bilinear.  Then, applying Algorithm \ref{algo} to compute the real delay margin embedded with this sampling / discretization couple, and comparing with the continuous-time based delay margin traditionally computed leads to Figure \ref{fig:exple01_margin}.

First, as a main remark, the delay margins computed by the \textbf{HSDMA} given in Algorithm \ref{algo}  and reported on Figure \ref{fig:exple01_margin} are exactly the same as the one obtained by an \ode resolution with variable step size, when introducing a transport delay in the loop. This then claims in favor to the proposed  \textbf{HSDMA} method, that provides the actual real delay margin. Then, as a second interesting remark, one may note that, unless the backward approach, in some cases (here on this example), the discretization may also result in an increase of delay margin, which is not obvious. This last comment may then be taken into account when selecting the discretization method. 

Moreover, as the discrete-time controller is approximated by a rational continuous-time one using the modified Loewner framework, even if the original controller is of dimension 2, in all cases computed, the resulting approximated controller is of considerably higher dimension, between 12 and 34 (depending on the method and sampling time).

\subsection{Comments on real-time and \ode solvers and real-time simulation}

For practitioners, the objective is usually to study the influence and stability issues of the discrete-time controller on the closed-loop system. In traditional scheme, this not easy to do and clearly, the standard delay margin is not precise enough to evaluate it when interconnecting continuous and discrete-time models. Therefore, the simulation-based approach is the last resort. 

Still, in the \ode framework, the delay margin computed must be interpreted as the sum of any delays  in the control-loop (representing computation time, transport delays, \etc) and the delays due to sample-hold blocks inevitably present when interconnecting systems with different sample times. Then, this delay margin can be verified by simulation by considering a closed loop system including a transport delay of $0.305s$ and two rate transition blocks, acting as Zero Order Hold. The complete sum of all these delays then leads to $0.3255$ seconds (when considering the specific consideration detailed above), which is exactly the one obtained with \textbf{HSDMA}.

%


In addition, the time-domain behavior analysis of the hybrid closed-loop system is also addressed thanks to a real-time execution framework. This framework is based on a toolchain including automatic code generation, scheduling tools and real-time execution environment. In our case, this toolchain has been exploited to increase the confidence level of our results previously presented. Note that part of this toolchain was successfully used in \cite{rosace_2014} for a complete  flight control laws case study.

For the stability margin of the considered system, we first discretize the plant with a period about a hundred times faster than the controller one's and we introduce a delay block in the loop. The corresponding multi-rate Simulink scheme is automatically translated into a synchronuous program with \cite{cocosimforge}, while preserving semantics of each \simulink blocks. Synchronuous language \code{LUSTRE} \cite{DBLP:journals/pieee/BenvenisteCEHGS03} is used for mono-periodic blocks whereas \code{PRELUDE} \cite{prelude} describes multi-period assemblies. \code{PRELUDE} is a formal language designed for the specification of the software architecture of a critical embedded control system and deals with real-time aspects of multi-periodic systems. In particular rate transition operators (source of delays in the execution) are automatically generated and enable the definition of communication patterns between blocks of different rates. The corresponding code is then simulate with the \code{SchedMcore} toolbox \cite{schedmcore11}.  

Then by modifying the delay's value and exploiting this toolchain, the time-domain behavior of the closed-loop system can be easily studied and the value of stability margin validated with respect of real-time constraints. We emphasize that the value of the delay leading to the limit of stability with real-time simulation corresponds to the one obtaining with \matlab \simulink, when rate transition blocks introduce minimal delays (no configuration options). In this case, safe and deterministic data transfers are not guaranteed, contrary to the \code{SchedMcore} toolbox. Moreover, and interestingly, the obtained delay margin exactly match the one obtained with the proposed \textbf{HSDMA}, which double check our proposed approach.


\section{Conclusions and perspectives}
\label{sec:conclusion}

In this paper we addressed the problem of evaluating the exact delay margin for hybrid system by focusing on the case where a discrete-time controller is interconnected to a continuous-time dynamical model. This problem is obviously not new, and rarely addressed as it. Even if we do not claim of having solved this problem, nevertheless,  in this paper we provide a simple but yet effective way to attack it by providing a systematic  procedure to evaluate some classical stability properties on a hybrid closed-loop system. This approach is based on a rational continuous-time model approximation of the discrete-time controller, connected to the standard continuous domain tools for delay stability margin computation. As a matter of fact, the resulting approach  is both efficient and numerically efficient (and scalable). Such an analysis might be considered for additional margin computation. 


\bibliographystyle{apalike}
\bibliography{biblio,_biblioCPV,_biblioPoussot}



\end{document}